
\input amstex
\documentstyle{amsppt}
\magnification 1200
\NoBlackBoxes
\NoRunningHeads
\def\1{\bold 1}

\def\Z{\Bbb Z}
\def\C{\Bbb C}

\def\N{\Bbb N}
\def\h{\frak h}
\def\d{\partial}

\def\gl{\frak{gl}_n}
\def\U{U_q\frak{gl}_n}

\def\Tr{\text{\rm Tr}}
\def\Id{\text{\rm Id}}
\def\lk{{\lambda+(k-1)\rho}}
\topmatter
\title Macdonald's polynomials and representations of quantum groups\endtitle
\author {\rm {\bf Pavel I. Etingof, Alexander A. Kirillov, Jr.} \linebreak
	\vskip .1in
   Department of Mathematics\linebreak
   Yale University\linebreak
   New Haven, CT 06520, USA\linebreak
   e-mail: etingof\@math.yale.edu,
			kirillov\@math.yale.edu}
\endauthor
\endtopmatter
\document
\centerline {December 11, 1993}
\vskip .1in
\heading{\bf Introduction}\endheading
\vskip .1in

Recently I.Macdonald defined a family of systems of orthogonal
symmetric polynomials depending of two parameters $q,k$
which interpolate between Schur's symmetric functions
and certain spherical functions on SL(n) over the real and p-adic
fields \cite{M}. These polynomials are labeled by dominant integral
weights of SL(n), and (as was shown by I.Macdonald) are uniquely defined by two
conditions: 1) they are orthogonal with respect to a certain weight
function, and 2) the matrix transforming them to Schur's symmetric
functions is strictly upper triangular with respect to the standard
partial ordering on weights (``strictly'' means that the diagonal entries
of this matrix are equal to 1). Another definition of Macdonald's
polynomials is that they are (properly normalized) common
eigenfunctions of a commutative set of $n$ self-adjoint partial difference
operators $M_1,...,M_n$ (Macdonald's operators) in the space of
symmetric polynomials.

In this paper we present a formula for Macdonald's polynomials
which arises from the representation theory of the quantum group
$U_q(\frak{sl}_n)$. This formula expresses Macdonald's
polynomials via (weighted) traces of intertwining operators
between certain modules over
$U_q(\frak{sl}_n)$.

The paper is organized as follows. In Section 1, we define Macdonald's
inner product, orthogonal polynomials, and commuting difference
operators, and compute the eigenvalues of these operators.
In Section 2, we review some facts about representations
of quantum groups that will be needed in the following sections.
In Section 3 we introduce weighted traces of intertwiners
(vector-valued characters) and prove an analogue of the Weyl
orthogonality theorem for them.
In Section 4 we formulate the main result -- the explicit formula for
Macdonald's polynomials for positive integer values of $k$
-- and give a complete proof of this formula. In Section 5, we
generalize the result of Section 4 to the case of an arbitrary $k$.
In Section 6, we construct Macdonald's operators from
the generators of the center of $U_q(\frak{sl}_n)$, and
derive an explicit formula for generic (non-symmetric) eigenfunctions
of Macdonald's operators using this construction.
\vskip .1in

{\bf Acknowledgements.} This paper was written at the time when
Ian Macdonald was giving a series of lectures
at Yale in November of 1993. We are grateful to Ian Macdonald
for attention to our work.

We would like thank our advisor Igor Frenkel
for systematic stimulating discussions. We are also grateful to
I.Cherednik, H.Garland, I.Grojnowski, and R.Howe for useful remarks.

The work of P.Etingof was supported by the Alfred P.Sloan foundation.
\vskip .1in

\heading {\bf 1. Macdonald's polynomials}\endheading
Here we give the definition and main properties of Macdonald's polynomials
for the root system of type $A_{n-1}$, following \cite{M}.

Let us fix $n\in \Z_+$. A sequence $\lambda=(\lambda_1\ldots\lambda_n)
\in\left(\Z_+\right)^n$ is called a partition if $\lambda_i\ge
\lambda_{i+1}$. We define a partial order on partitions: $\lambda>\mu$
if $\sum\lambda_i=\sum\mu_i$ and $\lambda_1=\mu_1,\ldots, \lambda_k=\mu_k,
\lambda_{k+1}>\mu_{k+1}$ for some $k<n$.

Let us consider polynomials of $n$ variables $x_1\ldots x_n$: $\Cal A=\C[x_1,
\ldots, x_n]$. For any
$\lambda \in\Z^n$, let $x^\lambda=x_1^{\lambda_1}\ldots
x_n^{\lambda_n}$. We have an obvious action of the Weyl group $S_n$ on
$\Cal A$.  We can take a basis of  $\Cal A^{S_n}$ formed by the orbitsums

$$m_\lambda = \sum\limits_{\mu\in S_n\lambda} x^\mu,\tag 1.1$$
where $\lambda$ runs through the set of all partitions. These functions
are orthogonal with respect to the inner product given by
$$<f,g>_0=[f\bar g]_0,$$
where
$$\bar g(x_1,\ldots,x_n)=g(x_1^{-1},\ldots,x_n^{-1}),$$
and $[\ ]_0\colon \C[x_1^{\pm 1},\ldots,x_n^{\pm 1}]\to \C$ is the
constant term:
$$\left[\sum a_\lambda x^\lambda\right]_0=a_0.$$

The main object of our study are Macdonald's polynomials, defined in
\cite{M}. This is a family of polynomials depending on two independent
variables $q,t$ and defined by the following theorem:
\proclaim{Theorem}{\rm (Macdonald)} There exists a unique family of
polynomials
$P_\lambda(x;q,t)\in \C(q,t)[x]$ $(x=(x_1,\ldots,x_n))$, where
$\lambda$  is a partition and
$\C(q,t)$ is the field of rational functions in $q,t$,
satisfying the following properties:

1. $P_\lambda(x;q,t)$ is symmetric under the action of $S_n$ on the
$x$'s.

2.
$P_\lambda(x;q,t)=m_\lambda(x)+\sum\limits_{\mu<\lambda}c_{\lambda\mu}
m_\mu(x)$

3. For fixed $q, t$ the polynomials $P_\lambda(x;q,t)$ are
orthogonal with respect to the inner product given by
$$<f,g>_{q,t}=[f\bar g\Delta_{q,t}]_0,\tag 1.2$$
where
$$\Delta_{q,t}(x)=\prod\limits_{i\ne j} \prod\limits_{m=0}^\infty
	\frac{1-q^{2m}x_ix_j^{-1}}{1-q^{2m}t^2 x_ix_j^{-1}}\tag 1.3 $$
\endproclaim

These polynomials are called Macdonald's polynomials (our notation differs
slightly from that of Macdonald: what we denote by $P_\lambda(x;q,t)$ in
the notations of \cite{M} would be $P_\lambda(q^2,t^2)$).

Also, often it is convenient to consider Macdonald's polynomials for
$t=q^k, k\in Z_+$; for example, for $k=0$ these polynomials reduce to the
orbitsums $m_\lambda$, and for $k=1$ to Schur's symmetric functions.
However, most of the properties of Macdonald's polynomials  obtained for
$t=q^k$ can be generalized to the
case when $q,t$ are independent variables.

In the future it will be convenient to use the following form of $\Delta$:

$$\Delta_{q,t}(x) =  \prod\limits_{\alpha\in R} \prod\limits_{m=0}^\infty
	\frac{1-q^{2m}x^{\alpha}}{1-q^{2m}t^2x^{\alpha}}
=\delta_{q,t}\bar \delta_{q,t}, \tag 1.4$$
where $R\subset\C^n$ is the root system of type $A_{n-1}$:
$R=\{\alpha_{ij}\}_ {i\ne j},
\alpha_{ij}=\varepsilon_i-\varepsilon_j,
\varepsilon_i=(0,\ldots,0,1,0,\ldots, 0)\in \Z^n $ (1 in the $i$-th
place), and

$$\delta_{q,t}(x)=	\prod\limits_{\alpha\in R^+} \prod\limits_{m=0}^\infty
	\frac{1-q^{2m}x^{\alpha}}{1-q^{2m}t^2x^{\alpha}}
\tag 1.5$$
$R^+=\{\alpha_{ij}\}_{i<j}$ being  the set of positive roots.

The proof of the theorem is based on the use of the following operator
$M_1$ in
the space $\C(q,t)[x_1,\ldots,x_n]^{S_n}$:
$$M_1=t^{1-n}\sum\limits_{i=1}^n\left(\prod\limits_{i\ne j}\frac{t^2x_i-
x_j} {x_i-x_j} \right)T_{q^2,x_i},\tag 1.6$$
where $(T_{q^2,x_i}f)(x_1,\ldots,x_n)=f(x_1,\ldots, q^2x_i,\ldots, x_n)$.
Then one can show that $M_1$ is self-adjoint with respect to the inner
product $<\cdot,\cdot>_{q,t}$ and that the eigenvalues of $M_1$ are distinct.
The Macdonald's polynomials defined above are just the eigenfunctions of
the operator $M_1$:
$$\gathered
M_1 P_\lambda(x;q,t)= c_\lambda^1 P_\lambda(x;q,t)\\
c_\lambda^1 =\sum\limits_{i=1}^nq^{2\lambda_i}t^{(n+1-2i)}
\endgathered\tag 1.7$$

Macdonald showed that the operator $M_1$ can be included in a
commutative family  of
difference operators (cf. \cite{Ch}). Namely, let
$$M_r=t^{r(r-n)} \sum\limits_{i_1<i_2<\ldots<i_r}
  \left(\prod\limits_
  {\gathered j\notin\{i_1\ldots i_r\}\\ l=1\ldots r\endgathered }
       \frac{t^2x_{i_l} -x_j}{x_{i_l}-x_j} \right)
	T_{q^2, x_{i_1}}\ldots T_{q^2,x_{i_r}}\tag 1.8$$

\proclaim{Proposition 1.1} {\rm (Macdonald)}

1.$ [M_i, M_j]=0$

2. $M_r$ is self-adjoint with respect to the inner product
$<\cdot,\cdot>_{q,t}$.

3. $M_rP_\lambda(x;q,t)=c_\lambda^r P_\lambda(x;q,t)$, where
$c_\lambda^r =\sum \limits_{i_1<\ldots<i_r}
\prod\limits_{l=1}^r q^{2\lambda_{i_l}}t^{(n+1-2i_l)} $.\endproclaim

\demo {Proof} Let us first prove that
$M_rm_\lambda=c_\lambda^rm_\lambda+\text{ lower order terms}$.
The proof is quite similar to that for $M_1$ (see \cite{M}).
Let us for simplicity assume that $t=q^k$. Introduce

$$\aligned
\delta(x)=&\prod\limits_{\alpha\in
R^+}(x^{\alpha/2}-x^{-\alpha/2})\\
=&(x_1\ldots x_n)^{\frac{1-n}{2}}\prod\limits_{i<j}(x_i-x_j)=
\sum\limits_{\sigma\in S_n} \epsilon(\sigma)x^{\sigma
\rho},\endaligned\tag 1.9$$
where $\rho=(\frac{n-1}{2},\frac{n-3}{2},\ldots,\frac{1-n}{2})$, and
$\epsilon(\sigma)$ is the sign of permutation $\sigma$. Also,
for brevity we write $I=\{i_1,\ldots,i_r\}$, where $i_1<\ldots<i_r$, and
$T_{q^2,x_I}=T_{q^2,x_{i_1}}\ldots T_{q^2,x_{i_r}}$. Then one can
easily check that

$$M_r=\sum\limits_{I: |I|=r}\delta^{-1}(T_{t^2,x_I}\delta)T_{q^2,x_I}=
\delta^{-1} \sum\limits_{\sigma}\epsilon(\sigma)x^{\sigma\rho}
t^{2\sum_{i\in I}(\sigma\rho)_i}T_{q^2,x_I}.$$
Now, if we denote  $S_{\lambda}=\{\sigma\in
S_n|\sigma(\lambda)=\lambda\}$, then $|S_\lambda|m_\lambda=\sum\limits
_{\sigma'\in S_n}x^{\sigma'\lambda}$, and therefore

$$|S_{\lambda}|M_rm_\lambda=\sum\limits_{I: |I|=r}\sum\limits_\sigma\sum
\limits_{\sigma'}
\delta^{-1}\epsilon(\sigma)x^{\sigma\rho}
t^{2\sum\limits_{i\in I}(\sigma\rho)_i}
q^{2\sum\limits_{i\in I} (\sigma'\lambda)_i} x^{\sigma'\lambda}.$$

Introducing $\sigma''$ such that $\sigma'=\sigma\sigma''$ we see that
$$|S_\lambda|M_rm_\lambda=\sum\limits_{\sigma''}\sum\limits_I
q^{2\sum\limits_{i\in I}(k\rho +\sigma''\lambda)_i}\chi_{\sigma''\lambda},$$
where
$$\chi_\mu=\delta^{-1}\sum\limits_{\sigma\in S_n}
\epsilon(\sigma)x^{\sigma(\mu+\rho)};\tag 1.10$$
if $\mu$ is a partition, $\chi_\mu$  is the Schur symmetric function, i.e. the
character of the corresponding finite-dimensional representation of
$\gl$. Obviously, $\chi_{\sigma''(\lambda)}=\sum\limits_{\mu\le
\lambda}a_\mu m_\mu$, and  non-zero
contribution to $m_\lambda$ is given by the terms with
$\sigma''\lambda=\lambda$. Therefore,
$$M_r m_\lambda=\left(\sum\limits_I q^{2\sum\limits_{i\in I}
(k\rho+\lambda)_i} \right)m_\lambda +\text{lower order terms}.$$

To complete the proof of the theorem it suffices to show that $M_r$ is
self-adjoint with respect to $<\cdot,\cdot>_{q,t}$, which is
straightforward. \qed\enddemo

\heading {\bf 2. The quantum group $\U$ and its representations. }
\endheading

Let $q$ be a formal variable. By definition (\cite{D1, J}), quantum
group $\U$ is an associative algebra with unit over the ring $\C(q)$
of rational functions in $q$ with generators $e_i,  f_i, i=1\ldots n-1$,
$q^{h_i}, i=1\ldots n$ and relations

$$ \gathered
[h_i,h_j]=0\\
[h_i, e_i]=e_i\qquad [h_i,f_i]=-f_i\\
[h_i,e_{i+1}]=-e_{i+1}\qquad [h_i,f_{i+1}]=f_{i+1}\\
[h_i,e_j]=[h_i,f_j]=0,\qquad j\ne i,i+1\\
[e_i,f_j]=\delta_{ij}\frac{q^{h_i-h_{i+1}}-q^{h_{i+1}-h_i}}{q-q^{-1}}
\endgathered\tag 2.1$$

$$ \gathered
e_i^2e_j-(q+q^{-1})e_ie_je_i +e_je_i^2=0, i=j\pm 1,\quad
f_i^2f_j -(q+q^{-1})f_if_jf_i +f_jf_i^2=0\\
[f_i,f_j]=0=[e_i,e_j],\quad |i-j|>1\endgathered\tag 2.2$$

It is known that $\U$ is a Hopf algebra with the following
comultiplication $\Delta$ and antipode $S$:

$$\gathered
\Delta e_i =e_i\otimes q^{(h_{i+1}-h_i)/2} +q^{(h_i-h_{i+1})/2}\otimes
e_i\\
\Delta f_i =f_i\otimes q^{(h_{i+1}-h_i)/2} +q^{(h_i-h_{i+1})/2}\otimes
f_i\\
\Delta h_i=h_i\otimes 1+1\otimes h_i\endgathered\tag 2.3$$

$$\gathered
S(e_i)=-e_iq^{-1}\\
S(f_i)=-f_i q\\
S(h_i)=-h_i\endgathered\tag 2.4$$

In the limit $q\to 1$, $\U$ becomes the universal enveloping algebra
of the Lie algebra $\gl$: one can identify the generators with the
matrix units as follows: $e_i=E_{ii+1}, f_i=E_{i+1 i}, h_i=E_{ii}$.

Like its classical analogue, $\U$ admits the following polarization:
$\U=U^-\cdot U^0\cdot U^+$, where $U^{\pm}$ is the subalgebra
generated by $e_i$ (respectively, $f_i$), and $U^0$ is the
algebra generated by $q^{h_i}$. $\U$ also admits an algebra automorphism
$\omega$ (Cartan involution), which transposes $U^+$ and
$U^-$:
$$
\omega e_i=-f_i,\quad
\omega f_i=-e_i,\quad
\omega h_i=-h_i \tag 2.5 $$
Note that $\omega$ is a coalgebra antiautomorphism.

Representation theory of $\U$ is quite parallel to the classical case.
Unless otherwise stated, we consider only finite-dimensional
representations. Define Cartan subalgebra $\h$ to be the linear span of
$h_i$; then every $\lambda=(\lambda_1,\ldots,\lambda_n)\in\C^n$ can be
considered  as a
weight, i.e. a an element of $\h^*$ by $\lambda(h_i)=\lambda_i$. We have a
bilinear form on the weights given by $<\lambda,\mu>=\sum\lambda_i
\mu_i$, which allows us to identify $\h\simeq \h^*$.
Define the set of integral weights $P=\{\lambda| \lambda_i-\lambda_j
\in\Z\}$ and the
set of dominant weights $P_{+}=\{\lambda|
\lambda_i-\lambda_{i+1}\in \Z_+\}$. Note that $\lambda \in P_{+}$ iff
$\lambda+ a(1,\ldots,1)$ is a partition for some $a\in\C$. We have
a natural order on $P$ which is defined precisely in the same way as in
Section 1. It is also  convenient to introduce
fundamental weights $\omega_i=(1,\ldots,1,0\ldots,0)$ ( $i$ ones),
 $i=1\ldots, n-1$. Then $\lambda\in P_{+}$ iff
$\lambda=a(1,\ldots, 1)+\sum n_i\omega_i, n_i\in \Z_+$.

We also introduce the root system $R=\{\varepsilon_i-\varepsilon_j,
i\ne j\}\subset \h^*$, positive roots
$R^+=\{\varepsilon_i-\varepsilon_j, i<j\}$ and simple roots
$\alpha_i=\varepsilon_i-\varepsilon_{i+1}, i=1,\ldots, n-1$, where
$\varepsilon_i$ is a basis in $\h^*$:
$\varepsilon_i(h_j)=\delta_{ij}$. Also, let $Q$ be the lattice in
$\h^*$ spanned by all roots, and $Q^+\subset Q$ be the semigroup
spanned by all positive roots.

For every $\lambda\in
P_{+}$ there
exists a unique irreducible finite-dimensional representation $V_\lambda$
generated by a highest weight vector $v_\lambda$ of weight $\lambda$. This
exhausts all irreducible finite-dimensional  representations of $\U$. The
characters (and therefore, the dimensions) of the representations
$V_\lambda$ are the same as for the classical case. Also, every
 representation is a direct sum of irreducibles.

We can define a tensor product of two representations by the formula
$\pi_{V\otimes W}(x) =(\pi_V\otimes\pi_W)\Delta(x)$. From complete
reducibility we know that $V_\lambda\otimes V_\mu\simeq \bigoplus N_{\lambda
\mu}^\nu V_\nu$, and the coefficients $N_{\lambda\mu}^\nu$ are the same as
 for $\gl$.

It is known that for any $V,W$ the representations $V\otimes W$ and
$W\otimes V$ are isomorphic, but the isomorphism is non-trivial. More
precisely (see \cite{D1}), there exists a universal R-matrix
$\Cal R\in \U\hat\otimes \U$ ($\hat\otimes$ should be understood as a completed
tensor product) such that
$$\check R_{V,W}= P\circ \pi_V\otimes \pi_W(\Cal R) \colon V\otimes W\to
W\otimes V
\tag 2.6$$
is an isomorphism of representations. Here $P$ is the transposition:
$Pv\otimes w=w\otimes v$. Also, it is known that $\Cal R$ has the following
form:
$$\gather
\Cal R=q^{-\sum h_i\otimes h_i}\Cal R^*, \quad \Cal R^*\in U^+\otimes  U^-\tag
2.7\\
(\epsilon\otimes 1)(\Cal R^*)=(1\otimes\epsilon)(\Cal R^*)=1\otimes
1,\endgather
$$
where $\epsilon\colon \U\to \C$ is the counit.
We will also use the following formula for $\check R^2$:
$$(\check R_{V_\lambda,V_\mu} \check R_{V_\mu,V_\lambda})|_{V_\nu\subset
V_\mu\otimes V_\lambda}= q^{c(\lambda)+c(\mu)-c(\nu)}\text{ Id}\tag 2.8$$
where
$c(\lambda)=<\lambda,\lambda+2\rho>,\quad
\rho=(\frac{n-1}{2},\frac{n-3}{2}, \ldots,\frac{1-n}{2})$.

We will also need the notion of dual representation. Namely, if $V$ is a
 representation of $\U$ then by definition $V^*$ is the
representation of $\U$ in the dual space to $V$ given by
$$<xv^*,v>=<v^*,S(x)v>.$$
One easily checks that in this case the canonical pairing $V^*\otimes V\to
\C$ and embedding $\C\to V\otimes V^*$ are homomorphisms of
representations (the order of factors is important here). Also, one has
canonical isomorphisms: $\text{Hom}_{\U}(V,W)=\text{Hom}_{\U}(\C, V^*\otimes
W)=(V^*\otimes W)^{\U}$. For an irreducible representation $V_\lambda$,
$\left(V_\lambda\right)^* \simeq\ V_{\lambda^*}$, where
$(\lambda_1\ldots\lambda_n)^* = (-\lambda_n,\ldots,-\lambda_1)$.
If we take $(V^*)^*$, we get another action of $\U$ on the same space
$V$. These two actions are isomorphic:
$q^{-2\rho}:V\to V^{**}$ is an isomorphism of $\U$-modules.
This is due to the fact that $S^2(x)=q^{-2\rho}xq^{2\rho}$.

Finally, if $V$ is an  irreducible representation of $\U$ let us
consider the action of
$\U$ in $V$ given by $\pi_{V^{\omega}}(x)=\pi(\omega x)$, where $\omega$
is the Cartan involution defined above. We denote $V$ endowed with this
action by $V^\omega$. One can easily check that $V\simeq (V^\omega)^*$
(which is, of course, equivalent to saying that $V^\omega\simeq V^*$); that
is, there exists non-degenerate pairing $V\otimes V^\omega\to \C$ which
commutes with the action of $\U$. Another way tp say it is to say that
there exists a non-degenerate bilinear pairing
(Shapovalov form)  $(\cdot,\cdot)_V\colon V\otimes V\to\C$ such that
$$(xv,v')_V=(v,\omega S(x)v')_V.\tag 2.9$$
This form is symmetric (which relies on $\omega S\omega = S^{-1}$).

Note also that $(V\otimes
W)^\omega=W^\omega\otimes V^\omega$ and that if $\Phi\colon V\to W$ is an
intertwiner then $\Phi$ is also an intertwiner considered as a map
$V^\omega\to W^\omega$.

\heading {\bf 3. Traces of intertwiners and the generalized Weyl
orthogonality  theorem}
\endheading
Let $V, U$ be finite-dimensional representations of $\U$, and
$\Phi\colon V\to V\otimes U$ be a non-zero intertwining operator for $\U$.
\proclaim{Definition} A vector-valued character is the following function
of $x=(x_1,\ldots,x_n)$:

$$\chi_\Phi(x_1,\ldots, x_n)=\Tr |_{V} (\Phi x_1^{h_1}\ldots
x_n^{h_n}).\tag 3.1$$
\endproclaim

 From the definition it is clear that $\chi_\Phi$ is a linear
combination of monomials $x^\mu$ where $\mu$ runs over the set of weights
of $V$. Thus, we can consider $\chi$ as an element of the group
algebra $\Cal A=\C(q)[P]\simeq\{\sum \limits_{\lambda\in P}a_\lambda
x^\lambda|a_\lambda\in \C(q),\text{ almost all }a_\lambda=0\}$. We
will sometimes  call
elements of $\Cal A$ generalized Laurent polynomials in $x_i$; also,
we will write $x^h$ instead of $x_1^{h_1}\ldots x_n^{h_n}$.
 Note that the elements of $\Cal A$  can also be
interpreted as functions on $\h$ by letting $x_i(\sum z_j
h_j)=e^{z_i}$. This is the same as considering the function on $\h$
given by $\chi(h)=\Tr|_V(\Phi e^h)$.

 In particular, for $V=V_\lambda$ $\chi_{\Phi}\in x^\lambda
\C(q)[\frac{x_2}{x_1},\ldots, \frac{x_n}{x_{n-1}}]\otimes U$; the
highest term  of $\chi$ is
$ux^\lambda$ and the lowest term is $u'x^{-\lambda^*}$ for some
$u,u'\in U$ . Note that in the
contrast with the classical case, $\chi_\Phi$ is not $S_n$ symmetric
if $U$ is not a trivial representation.

Using the notion of dual representation, we can rewrite $\chi_\Phi$ as
follows:
we can identify $\Phi$ with an intertwiner $\Phi\colon V^*\otimes V\to U$;
then $\chi_\Phi(x)=\Phi(1\otimes x^h)\sum v^i\otimes v_i$, where $v_i, v^i$
are the dual bases in $V,V^*$. Note that $\sum v^i\otimes
v_i=(1\otimes q^{-2\rho})\1$, where $\1=i(1), i\colon \C\to V^*\otimes
V$ being an embedding of $\U$-modules.

In particular, this implies that if we substitute  $x_i=q^{2\rho_i}$
in $\chi$ then $\chi(q^{2\rho})=0$
if $U$ is a non-trivial irreducible  representation.

The space of
all intertwiners $\Phi\colon V_\lambda\to V_\lambda\otimes U$ is
isomorphic to the
space $(V_\lambda^*\otimes V_\lambda\otimes U)^{\U}$. Let us
choose a basis $\Phi_{\lambda, i}$ in
this subspace, orthonormal with respect to Shapovalov form.
This gives us the basis in the space of $U$-valued
characters.

\proclaim{Generalized Weyl Orthogonality Theorem} The characters
$\chi_{\lambda,i}$ are orthogonal with respect to the following inner
product:
$<f,g>_1=[(f,\bar g)_U \Delta]_0$,
where $\Delta=\prod\limits_{\alpha\in R}(1-x^\alpha)$,
$(\cdot,\cdot)_U$ is the Shapovalov form   and all the other
notations are as in Section 1.
\endproclaim
\demo{Proof}
Let $\chi_1=\Tr|_{V_\lambda} (\Phi_\lambda x^h)$,
$\chi_2=\Tr|_{V_\mu} (\Phi_\mu x^h)$. As was explained above, we can as
well consider $\Phi_\mu$ as an intertwiner $V_\mu^\omega\to
U^\omega\otimes V_\mu^\omega$. Thus,
$(\chi_1(x),\chi_2(x^{-1}))_U=\Tr|_{V_\lambda\otimes V_\mu^\omega}
(\Psi x^h\otimes x^h)$ (note the change of sign of $h$ in the second
factor!),  where the intertwiner $\Psi\colon
V_\lambda\otimes V_\mu^\omega\to V_\lambda\otimes V_\mu^\omega$ is
defined as the following composition

$$V_\lambda\otimes V_\mu^\omega @>{\Phi_\lambda\otimes \Phi_\mu^\omega}>>
 V_\lambda\otimes U\otimes
U^\omega\otimes V_\mu^\omega @>{\text{Id}\otimes
(\cdot,\cdot)_U\otimes \text{Id}}>>
V_\lambda\otimes V_\mu^\omega.$$

Since $V_\lambda\otimes V_\mu^\omega=\bigoplus N_\nu V_\nu$, we see that
$(\chi_1(x),\chi_2(x^{-1}))_U$ is a linear combination
of usual characters $\chi_\nu(x)$.
But since these characters are the same as
for $\gl$, we know that $[\chi_\nu(x)\Delta]_0=0$ unless $\nu=0$.
On the other hand, it is known that if $\lambda\ne \nu$ then the
decomposition of $V_\lambda\otimes V_\nu^\omega$ does not contain
the trivial representation (i.e. $N_0=0$); thus, in this case $\chi_1$
and $\chi_2$ are orthogonal.

If $\lambda=\nu$ then $\chi_{\Phi_i}$ are pairwise orthogonal by definition.
\qed

\enddemo

\heading {\bf 4. The main theorem}\endheading
Through this section, we assume $k\in \N$ and show how one gets
Macdonald's polynomials $P_\lambda(x;q,q^k)$ as vector-valued characters.
Let $U$ be the finite-dimensional representation
of $\U$ with the highest
weight $(k-1)n\omega_1-(k-1)(1,\ldots,1)=(k-1)(n-1, -1,\ldots,-1)$; as
a $U_q\frak {sl} _n$-module, this is a $q$-analogue of the representation
$S^{(k-1)n}\C^n$.
Note that  all the weight subspaces  in $U$ are
one-dimensional; this property will be very useful to us.

\proclaim{Lemma 1} A $\U$-homomorphism $\Phi\colon  V_\lambda\to
V_\lambda\otimes U$ exists iff $\lambda-(k-1)\rho\in P_{+}$; if it
exists,  it is unique
up to a factor.\endproclaim

As we discussed before, it suffices to prove this lemma for $\gl$, which
is a standard exercise. Therefore, let us consider the (non-zero) intertwiners

$$\Phi_\lambda: V_{\lambda+(k-1)\rho} \to
V_{\lambda+(k-1)\rho}\otimes U
,\qquad \lambda\in P_{+},\tag 4.1$$
and the corresponding traces
$$\varphi_\lambda(x)=\chi_{\Phi_\lambda}(x)=
\Tr|_{V_{\lambda+(k-1)\rho}}(\Phi_\lambda x^h);\tag 4.2$$

As we discussed before, $\varphi_\lambda(x)$ has the form
$\varphi_\lambda(x)=x^{\lk} p(x), p(x)\in
\C(q)[\frac{x_2}{x_1},\ldots,
\frac{x_n}{x_{n-1}}]$.
It takes
values in the zero-weight subspace $U[0]$, which is one-dimensional;
therefore, we can consider it as a complex-valued function. We choose
the normalization of $\Phi$ and the identification $U[0]\simeq\C$ in
such a way that

$$\varphi_\lambda (x) = x^{\lk}+ \ldots \tag 4.3$$

\proclaim{Proposition 4.1} $\varphi_\lambda(x)$ is divisible by
$\varphi_0(x)$, and the ratio is a  symmetric generalized Laurent
polynomial in  $x_i$ with highest term $x^\lambda$.\endproclaim
\demo{Proof}
Let $\lambda=a(1,\ldots 1)+ \sum n_i\omega_i$. We prove the statement
by induction in height of $\lambda$: $|\lambda|=\sum n_i$. If the
height is zero, this is obvious. Suppose we've proved the statement
for all $\lambda$ of height $\le N$; let us prove it for $\lambda'$ of
height $N+1$. Write $\lambda'=\lambda+\omega_i$ for some $\lambda$ of
height $N$. Then

$$V_{\omega_i}\otimes V_\lambda =V_{\lambda'} + \sum \limits_{|\nu|\le
N}C_\nu V_\nu.$$

By induction assumption, it suffices to prove that $\Tr(\Phi x^h)$
is divisible by $\varphi_0(x)$ and the ratio is symmetric,
where $\Phi\colon V_{\omega_i}\otimes V_\lambda\to V_{\omega_i}\otimes
V_\lambda\otimes U$ is such that its restriction to $V_{\lambda'}$ is
not zero. Let
us take $\Phi=\text{Id}\otimes \Phi_\lambda$. Then $\Tr(\Phi
x^h)=\chi_{\omega_i}(x) \varphi_\lambda(x)$. Since $\chi_{\omega_i}$
is a symmetric Laurent polynomial in $x_i$, and by induction assumption
the statement of the proposition holds for $\varphi_\lambda$,
we get the
desired result. \qed\enddemo

Now we can formulate our main theorem:

\proclaim{Theorem 1} If $\lambda$ is a partition then
$\frac{\varphi_\lambda(x) }{\varphi_0(x)}$ is the Macdonald's polynomial
$P_\lambda(x;q,q^k)$.\endproclaim

\demo{Proof} We have shown that $\varphi_\lambda(x)/\varphi_0(x)$ are
symmetric Laurent polynomials with highest term $x^\lambda$. It
implies that $\varphi_\lambda(x)/\varphi_0(x)\in
\C[x_1,\ldots,x_n]^{S_n}$.   Thus, it
suffices to prove that they are orthogonal with respect to the inner
product $<\cdot,\cdot>_{q,t}$ given by (1.2). This follows from
the generalized Weyl orthogonality theorem
and the following lemma:
\proclaim{Main Lemma}
$$\varphi_0(x)=q^{\frac{n(n-1)k(k-1)}{4}}\prod\limits_{i=1}^{k-1}\prod
\limits_{\alpha\in R^+}
(q^{-i}x^{\alpha/2}-q^ix^{-\alpha/2})\tag 4.4$$\endproclaim

Let us first show how to deduce the statement of the theorem from this
lemma. Indeed, we know from generalized Weyl orthogonality theorem that
$[\varphi_\lambda \bar \varphi_\mu\Delta]_0=0$ if $\lambda\ne \mu$. Therefore,

$$\left[ \frac{\varphi_\lambda}{\varphi_0}
\overline{\left(\frac{\varphi_\mu}{\varphi_0}\right)}
\varphi_0\bar\varphi_0\Delta\right]_0=0.\tag 4.5$$

Since
$$\varphi_0\bar\varphi_0\Delta=\prod\limits_{\alpha\in\
R}\prod\limits_{i=0}^{k-1} (1-q^{2i}x^{\alpha})=\Delta_{q,t}(x),$$
(4.5) is just the condition of orthogonality with respect to the inner
product $<\cdot, \cdot>_{q,t}$.

Now let us prove the Main Lemma. Unfortunately, straightforward computation
encounters serious difficulties; thus, we have to find a bypass.
First, we prove the following statement:

\proclaim{Lemma 1} $\varphi_\lambda \bar \varphi_\lambda$ is
$S_n$-symmetric\endproclaim

This follows from the construction used in the proof of the
generalized Weyl orthogonality theorem, where we proved that
$\varphi_\lambda \bar \varphi_\lambda$ is a linear combination of
characters of irreducible representations.

Now comes the key statement:

\proclaim{Lemma 2} $\varphi_0(x)$ is divisible by
$(1-q^{2j}x^{-\alpha_i})$ for any simple root $\alpha_i$ and $1\le
j\le k-1$.\endproclaim

To prove it, recall that $\varphi_0(x)=\Tr (\Phi_0 x^h)$, where
$\Phi_0\colon V_{(k-1)\rho}\to V_{(k-1)\rho}\otimes U$. Fix a simple
root $\alpha_i$ and consider $F_i=f_iq^{(h_{i+1}-h_i)/2}$. Then
$$\Delta(F_i)= F_i\otimes q^{h_{i+1}-h_i}+1\otimes F_i\tag 4.6$$

Consider $\Tr(\Phi_0 F_i^jx^h)\in U[-j\alpha_i]$. Then, using
the fact that $\Phi$ is an intertwiner and  (4.6), we see that

$$\Tr(\Phi_0 F_i^jx^h)=q^{2j}\Tr((F_i\otimes 1)\Phi_0 F_i^{j-1}x^h)
+F_i\Tr (\Phi_0 F_i^{j-1}x^h)$$

Now we can use the cyclic property of trace and relation $x^hF_i=F_i x^h
\frac {x_{i+1}}{x_i}$ to get

$$\Tr(\Phi_0 F_i^jx^h)=q^{2j}\frac{x_{i+1}}{x_i}\Tr (\Phi_0 F_i^j x^h)
+F_i\Tr (\Phi_0 F_i^{j-1}x^h)$$

Repeating this process, we see that
$$\Tr(\Phi_0 F_i^jx^h)=\prod\limits_{m=1}^j
\frac{1}{1-q^{2m}\frac{x_{i+1}}{x_i}} F_i^j\Tr(\Phi_0x^h)$$

But we know that $F_i^j|_{U[0]}\ne 0$ for $j\le k-1$. Also, it is easy
to see that the left-hand side is a Laurent polynomial in $x_i$. This
proves Lemma 2. \qed

Now, let us consider $\varphi_0\bar \varphi_0$. We know that it is divisible
by $1-q^{2j}x^{\alpha_i}$ for any simple root $\alpha_i$. But we
also know that it is $S_n$-invariant. Since the Weyl group acts
transitively on the root system (in the simply-laced case), we see
that $\varphi_0\bar \varphi_0$ is divisible by $1-q^{2j}x^{\alpha}$ for
any $\alpha\in R$. Comparing the degrees in each variable, we see that

$$\varphi_0\bar\varphi_0=c\prod\limits_{j=1}^{k-1}\prod\limits_{\alpha\in R}
(1-q^{2j}x^{\alpha})$$
for some constant $c$ depending on $q,k$.

This is only possible if
$$\varphi_0=\pm  q^{N}\prod\limits_{j=1}^{k-1}
\prod\limits_{\alpha\in R'}
(q^{-j}x^{\alpha/2}-q^{j}x^{-\alpha/2})\tag 4.7$$
for some subset $R'\subset R$ such that $R=R'\sqcup -R'$
(normalization of $\varphi_0$ uniquely determines the factor in front of
the product). We want to
prove $R'=R^+$. To do it, let us compare the lowest terms on both
sides of (4.7). The lowest term of the right hand side is

$$\prod\limits_{\alpha\in R'} (-1)^{k-1} q^{\pm k(k-1)}x^{-\lambda_0},$$
where one takes $+$ sign if $\alpha\in R^+$ and $-$ otherwise,
$\lambda_0=(k-1)\rho$ (if we normalize so that the highest term is
$x^{\lambda_0}$). Thus, to complete the proof of the theorem it
suffices to prove the following lemma:

\proclaim{Lemma 3}
$$\varphi_0(x)=x^{\lambda_0}+\ldots \pm
q^{\frac{k(k-1)n(n-1)}{2}} x^{-\lambda_0},\tag 4.8$$
where $\lambda_0=(k-1)\rho$.
\endproclaim

This again involves several steps, which we briefly outline. First,
note that $V_{\lambda_0}\simeq V_{\lambda_0}^*$. Similar to the discussion at
the end of Section 2 and beginning of Section 3, we can rewrite $\varphi$
as follows:
$$\varphi_0(x)= \Phi (1\otimes q^{-2\rho} x^h)\1 ,\tag 4.9$$
where $\1=i(1)$, $i\colon \C\to V_{\lambda_0}\otimes V_{\lambda_0}$,
$\Phi\colon V_{\lambda_0} \otimes V_{\lambda_0}\to U$ are non-zero
homomorphisms of $\U$-modules (both $i$ and $\Phi$ exist and are
unique up to a factor).

Let us write
$\1=v_{\lambda_0}\otimes v_{-\lambda_0}+\ldots +
av_{-\lambda_0}\otimes v_{\lambda_0},$ for
some $v_{\lambda_0}\in V_{\lambda_0}[\lambda_0],
v_{-\lambda_0}\in V_{\lambda_0}[- \lambda_0]$ and $a\in \C(q)$.

Now, consider the intertwiner $\check R=\check
R_{V_{\lambda_0},V_{\lambda_0}}  \colon
V_{\lambda_0}\otimes V_{\lambda_0}\to V_{\lambda_0}\otimes V_{\lambda_0}$,
defined in Section 2.
Let us write $V_{\lambda_0}\otimes V_{\lambda_0}=\bigoplus C_\mu V_\mu$;
in particular, the trivial representation $\C$
occurs in this decomposition with multiplicity 1. Then formula
(2.8) for $\check R^2$ implies

$$\check R|\1=\pm q^{c(\lambda_0)}\1,$$
where $c(\lambda)=<\lambda, \lambda+2\rho>$. On the other hand, we
know (from formula 2.7) that $\check R(v_{\lambda_0} \otimes
v_{-\lambda_0})= q^{<\lambda_0,\lambda_0>}v_{-\lambda_0}\otimes
v_{\lambda_0}$. This implies

$$\1=v_{\lambda_0}\otimes v_{-\lambda_0}+\ldots \pm
q^{c(\lambda_0)-<\lambda_0,\lambda_0>}v_{-\lambda_0}\otimes
v_{\lambda_0}=v_{\lambda_0}\otimes
v_{-\lambda_0}+\ldots +q^{<\lambda_0,
2\rho>}v_{-\lambda_0}\otimes v_{\lambda_0}$$

Denote  $\Phi(v_{\lambda_0}\otimes
v_{-\lambda_0})=u_0\in U[0]$. Let us find  $\Phi(v_{-\lambda_0}\otimes
v_{\lambda_0})$. Again, since $U$ occurs in the decomposition of
$V_{\lambda_0}\otimes V_{\lambda_0}$ with multiplicity one, we have

$$\check R|_{U\subset V_{\lambda_0}\otimes
V_{\lambda_0}}=\pm q^{c(\lambda_0)-\frac{c(\nu)}{2}},$$
where $c(\lambda)=<\lambda, \lambda+2\rho>$ and
$\nu=(k-1)(n-1,-1,\ldots, -1)$ is the highest weight of $U$. Thus,
$\Phi \check R=\pm q^{c(\lambda_0)-\frac{c(\nu)}{2}}\Phi$. Considering
$\Phi\check R (v_{\lambda_0}\otimes v_{-\lambda_0})$, we find that

$$\pm q^{c(\lambda_0)-\frac{c(\nu)}{2}}\Phi(v_{\lambda_0}\otimes
v_{-\lambda_0})= \pm q^{<\lambda_0,\lambda_0>}\Phi
(v_{-\lambda_0}\otimes v_{\lambda_0})$$

Thus, if choose identification $U[0]\simeq \C$ so that
$\Phi(v_{-\lambda_0}\otimes v_{\lambda_0})=1$ then

$$\aligned
\varphi_0(x)=&\Phi(1\otimes q^{-2\rho}x^h)\1\\
 =&\Phi(1\otimes q^{-2\rho}x^h)(v_{\lambda_0}\otimes v_{-\lambda_0}+\ldots +
	q^{<\lambda_0, 2\rho>}v_{-\lambda_0}\otimes v_{\lambda_0})\\
= &x^{\lambda_0}+\ldots\pm q^{\frac{c(\nu)}{2}} x^{-\lambda_0 }
\endaligned
 \tag 4.10 $$

Since $c(\nu)=k(k-1)n(n-1)$, we get the statement of Lemma 3.
This completes the proof of the theorem.
\qed
\enddemo

\heading {\bf 5. The case of generic $k$} \endheading

In this section we show how to get Macdonald's polynomials
for the case when $q$ and $t$ are independent variables.
However, it will be convenient to
introduce formal variable $k$ such that $t=q^k$; thus, $q$ and $q^k$ are
algebraically independent variables. One can check that all the formulas
can be rewritten in such a way that $k$ appears only in the expression
$q^k$ and thus we could avoid using $k$, writing everything entirely in
terms of $q,t$; however, this would make our construction less
transparent. Also, we must consider the algebra $\U$, as well as the
representations, over the field $\C(q,t)$ rather than $\C(q)$.

Let $M_\mu$ be a Verma module with highest weight $\mu$ over $\U$,
$v_\mu$ be the corresponding highest weight vector. We choose a
homogeneous basis $a_i$ in $U^-$; then the basis in $M_\mu$ is given
by $a_iv_\mu$. In particular, this applies to the module
$M_{\lambda+(k-1)\rho}$, which is a natural generalization of the
module considered in the previous section. Note that if $k$ is a
formal variable then this module is irreducible.

We can also introduce the analogue of the module $U$. Indeed, let

$$W_{k}=\{(x_1\ldots x_n)^{k-1} p(x), p(x)\in \C[x_1^{\pm 1},\ldots,
x_n^{\pm 1}], \text{deg }p=0\}$$
with the action of $\U$ given by
$$\gathered
h_i\mapsto x_i\frac{\d}{\d x_i}-(k-1)\\
e_i\mapsto x_i D_{i+1}\\
f_i\mapsto x_{i+1} D_i\\
(D_i f)(x_1,\ldots, x_n)=\frac{f(x_1,\ldots, qx_i,\ldots ,x_n)-
f(x_1,\ldots, q^{-1}x_i,\ldots ,x_n)}{(q-q^{-1})x_i}
\endgathered\tag 5.1$$

$W_k$ is an irreducible infinite-dimensional module over $\U$. The set of
weights of $W_k$ is the root lattice $Q$, and every
weight subspace is one-dimensional:

$$W_k[\lambda]=\C w_\lambda, \qquad w_\lambda=(x_1\ldots x_n)^{k-1}
x^\lambda.$$

If we replace formal variable $k$ in the formulas above by a positive
integer $k$
then $W_k$ has a finite-dimensional submodule
$U_k=W_k\cap \C[x_1,\ldots, x_n]$; it coincides with the module
$U$ defined in Section 4.

\proclaim{Lemma 5.1} For every $\lambda\in P_{+}$ there exists a
unique up to a constant factor  intertwiner
$$\tilde \Phi_\lambda^k\colon M_{\lambda+(k-1)\rho}\to M_{\lambda
+(k-1)\rho}\otimes W_k.\tag 5.2 $$\endproclaim

We use the notation $\tilde\Phi$ to distinguish these intertwiners
from those for finite-dimensional modules introduced in Section 4; the
same convention applies to all other notations.

Proof is based on the general fact: if Verma module $M_\mu$  is irreducible
then the space of intertwiners $M_\mu\to M_\mu\otimes W$ is in one-to-one
correspondence with the zero-weight subspace $W[0]$.

Let us fix the normalization of $\tilde \Phi_\lambda^k$ by
choosing a highest-weight vector
$v_{\lk}\in V_{\lambda+(k-1)\rho}[\lambda+(k-1)\rho]$ and requiring
that $\tilde\Phi_\lambda^kv_{\lk} =v_{\lk} \otimes w_0 +\ldots$.
Then one can find explicit formulas for matrix elements of  $\tilde \Phi$
 as follows: write

$$\tilde\Phi_\lambda^k(a_i v_{\lk})=\sum \tilde R_{\lambda}^{ijl}
a_jv_{\lk}\otimes w_l\tag 5.3$$

Then the condition for $\tilde\Phi$ to be an intertwiner can be
rewritten as a system of linear equations on $\tilde R_\lambda^{ijl}$. Due to
Lemma 5.1, this system has a unique solution. From this approach one
can easily see that $\tilde R_\lambda^{ijl}$ is a rational function in
$q, q^k$.

Similar to Section 4, define the trace:

$$\tilde \varphi_\lambda^k(x)=\Tr|_{M_{\lambda+(k-1)\rho}}(\tilde
\Phi_\lambda^k x^h) \tag 5.4$$

Again, $\tilde \varphi_\lambda^k(x)$ takes values in $W_k[0]$, which
is one-dimensional; so we consider $\tilde \varphi$ as a scalar-valued
function, identifying $W_{k}[0]\simeq \C$ so that $w_0\mapsto 1$. Then

$$\tilde
\varphi_\lambda^k(x)=x^{\lambda+(k-1)\rho}(1+\sum\limits_{\mu\in
Q_+}\tilde R_{\lambda \mu}(q,q^k)x^{-\mu}),\tag 5.5$$
and $\tilde R_{\lambda \mu}$ are rational functions of $q,q^k$.

\proclaim{Theorem 2}
$\frac{\tilde\varphi_\lambda^k(x)}{\tilde\varphi_0^k(x)}$ is the Macdonald's
polynomial $P_\lambda(x; q,q^k)$\endproclaim

\demo{Proof} Let us recall the traces considered in Section 4. Let
$k$ be a positive integer, $\lambda\in P_{+}$. Then we have defined

$$\varphi_\lambda^k=\Tr |_{V_\lk}(\Phi_\lambda^k x^h),\tag 5.6$$
where

$$\Phi_\lambda^k\colon V_\lk\to V_\lk\otimes U\tag 5.7$$
is an intertwiner. First, note that $U$ is a submodule in the module
$W_k$ defined in the beginning of this section, so we can as well
consider $\Phi$ as an intertwiner $V_\lk\to V_\lk\otimes W_k$. Next,
the irreducible module $V_\lk$ is a
factormodule of Verma module $M_\lk$. Moreover, if

$$\mu=\sum n_i \alpha_i, \quad n_i\in\Z_+, \quad \sum n_i<k\tag 5.8$$
then $\text{dim }V_{\lambda+(k-1)\rho}[\lambda+(k-1)\rho-\mu]
=\text{dim }M_{\lambda+(k-1)\rho}[\lambda+(k-1)\rho-\mu]$. Thus, if we
consider elements $a_i$ of the  basis in $U^-$ such that $\mu=-\text{weight
}a_i$ satisfies condition (5.8) then the vectors $a_i v_\lk$ form a
basis in the corresponding weight subspaces of $V_\lk$.
Let us consider the restriction of the operator $\Phi$ to these
subspaces. Then it can be written in the form

$$\Phi_\lambda^k (a_i v_\lk)=\sum R^{ijl; k}_\lambda a_j
v_\lk\otimes w_l\tag 5.9$$

The coefficients $R^{ijl; k}_\lambda(q)$  are rational functions of $q$.
They can be found by solving the system of equations expressing the
intertwining property of $\Phi$. This is the same system which defined
the coefficients $\tilde R^{ijl}_\lambda(q, q^k)$ in the expansion
(5.3) of the intertwiners $\tilde \Phi$, but now we consider
$k$ as a positive integer, not a formal variable. Still one can check
that if we restrict ourselves to considering only $R^{ijl;k}_\lambda$
such that both $-\text{wt }a_i, -\text{wt }a_j$ satisfy (5.8) then
this system has a unique solution. Thus, we have the following lemma.
\proclaim{Lemma} For fixed $\lambda, i,j,l$ such that $-\text{weight
}a_i, -\text{weight }a_j$ satisfy (5.8),

$$R^{ijl;k}_\lambda(q)=\tilde R^{ijl}_\lambda(q,q^k)\tag 5.10$$
for $k\in \Z_+, k>>0$. Here the right-hand side should be understood
as a rational function of $q$ obtained by substituting $t=q^k,
k\in\Z_+$ in the rational function of two variables $\tilde
R^{ijl}_\lambda(q,t)$. \endproclaim

\proclaim{Corollary 1} If we write

$$\varphi_\lambda^k(x)=x^{\lk}\bigl(1+\sum\limits_{\mu\in
Q_+}R_{\lambda\mu}^k(q) x^{-\mu}\bigr)\tag 5.11$$
then for fixed $\lambda, \mu$

$$R^k_{\lambda\mu}(q)=\tilde R_{\lambda\mu}(q, q^k)$$
for $k\in \Z_+, k>>0$. \endproclaim

Let us consider the ratio $\tilde \varphi_\lambda^k/\tilde\varphi_0^k$.
Clearly, it can be written in the form

$$\frac{\tilde
\varphi_\lambda^k(x)}{\tilde\varphi_0^k(x)}=x^{\lambda}
\bigl(1+\sum\limits_{\mu\in Q_+}\tilde
Q_{\lambda\mu}(q,q^k)x^{-\mu}\bigr) \tag 5.12$$

Similarly,
$$\frac{\varphi_\lambda^k(x)}{\varphi_0^k(x)}=x^{\lambda}
\bigl(1+\sum\limits_{\mu\in Q_+}Q_{\lambda\mu}^k(q)x^{-\mu}\bigr)
 \tag 5.13$$
(in fact, the latter sum is finite due to Theorem 1).

Then Corollary 1 above
immediately implies the following:
\proclaim{Corollary 2} For fixed $\lambda, \mu$,

$$Q_{\lambda\mu}^k(q)=\tilde Q_{\lambda\mu}(q,q^k)\tag 5.14$$
for $k\in \Z_+, k>>0$.
\endproclaim

On the other hand, Theorem 1 in the previous section claims that if
one writes Macdonald's polynomials in the form

$$P_\lambda(x;q,t)=x^{\lambda}\bigl(1+\sum\limits_{\mu\in
Q_+}P_{\lambda\mu}(q,t)x^{-\mu}\bigr) $$

then
$$Q_{\lambda\mu}^k(q)=P_{\lambda\mu}(q,q^k)\qquad \text{ for all
}k\in\N\tag 5.15$$

Comparing (5.15) and (5.14), we see that

$$\tilde Q_{\lambda\mu}(q,q^k)=P_{\lambda\mu}(q,q^k)\qquad \text{ if }
k\in\Z_+, k>>0.$$
But this is possible only if $P_{\lambda\mu}=\tilde Q_{\lambda\mu}$ as
functions of two varibles $q,t$. Thus, the ratio (5.12) equals to the
Macdonald's polynomial $P_\lambda(x;q, t)$. \qed
\enddemo

\heading{\bf 6. The center of $\U$ and Macdonald's operators}\endheading
In this section we show how one can get Macdonald's operators $M_r$
introduced in Section 1 from the quantum group $\U$. This construction is
parallel to the one for $q=1$ (see \cite{E}).

For simplicity, in this section we assume that $t=q^k, k\in \N$.
Consider functions $f$ of $n$ variables $x_1,\ldots, x_n$ and introduce
the ring of difference operators, acting on these functions:

$$\text{DO }=\{ D =\sum\limits_{\alpha\in \Z^n} a_\alpha T_\alpha|\text{
almost all }a_\alpha=0\},\tag 6.1$$
where $(T_\alpha f)(x_1,\ldots, x_n)=f(q^{\alpha_1}x_1,\ldots,
q^{\alpha_n}x_n)$, and $a_\alpha$ are rational functions in $x_i,q$ with
poles only at the points where $x^\mu q^m=1$ for some $\mu\in \Z^n, m\in
\Z$.

As before, let us consider a non-zero intertwiner $\Phi\colon V\to
V\otimes W$, where $V$ is a highest-weight module over $\U$ and $W$ is
an arbitrary module with finite-dimensional weight spaces ($V,W$ need
not be finite-dimensional), and define the corresponding trace
$\varphi(x)=\Tr|_V(\Phi x^h)$.  This function takes values
in $W$.

\proclaim{Theorem 3} For any $u\in \U$ there exists a difference
operator $D_u\in \text{DO}\otimes \U$, independent of the choice of $V,W$
and the intertwiner $\Phi$ such that

$$\Tr|_V(\Phi u x^h)=D_u\Tr|_V(\Phi x^h).\tag 6.2$$

$D_u$ is defined uniquely modulo the left ideal in $\U$ generated by
$q^{h_i}-1$; thus, $D_u f$ is well defined for any function
$f(x_1,\ldots, x_n)$ with values in $W[0]$ .

\endproclaim

\demo{Proof}

Let us assume that $u$ is homogeneous:
 $u\in \U[\mu]$. Without loss of generality we can assume that
$u$ is a monomial in the generators
$e_i, f_i, q^{h_i}$ of the form
$u=u^- u^0 u^+$, $u^\pm\in U^\pm, u^0\in U^0$. Define $\text{sdeg
}u=\text{deg }u^+ -\text{deg }u^-$, where $\text{deg }e_i=-\text{deg
}f_i=1$. We prove the theorem by induction in $\text{sdeg }u$.

If $\text{sdeg }u=0$ then $u=u^0=q^{\sum \alpha_i h_i}$ for some
$\alpha\in\Z^n$. Then it follows immediately from the definition that
$D_u=T_\alpha$, so the theorem holds.

Let us make the induction step. Since $\Phi$ is an intertwiner, $\Tr(\Phi
u x^h)=\Tr (\Delta(u)\Phi x^h)$. From the definition of comultiplication
one easily sees that

$$\Delta (u)= u\otimes q^{\sum \alpha_i h_i}+ \sum u'_j\otimes v_j$$
for some $\alpha\in \Z^n$, and  $\text{sdeg }u'_j<\text{sdeg }u$. Thus,

$$\Tr(\Phi u x^h)= q^{\sum \alpha_i h_i} \Tr (\Phi x^h u )+ \sum
v_j\Tr(\Phi x^h u_j).$$

Since commuting with $x^h$ does not change $\text{sdeg }u_j$, by
induction assumption we can write

$$\Tr(\Phi u x^h)= q^{\sum \alpha_i h_i} \Tr (\Phi x^h u )+
 D'\Tr(\Phi x^h)$$
for some $D'\in \text{DO}\otimes \U$. Since $u\in \U[\mu]$,
$x^hu=x^\mu ux^h$, and $\Tr(\Phi x^h u)\in W[\mu]$, so
$$\Tr(\Phi u x^h)= q^{<\alpha,\mu>}x^\mu \Tr (\Phi u x^h )+
 D'\Tr(\Phi x^h),$$
and thus,
$$\Tr(\Phi u x^h)=\frac{1}{1- q^{<\alpha,\mu>}x^\mu} D'\Tr(\Phi
x^h). $$

This proves the existence part of the theorem. Uniqueness follows from the
following lemma:
\proclaim{Lemma} Let us fix a $\U$-module $W$ with finite-dimensional
weight spaces. If
$D\in \text{DO}\otimes \text{Hom  }(W[0], W[\mu])$ is such
that
$$D\varphi =0$$
for any $\varphi(x)=\Tr (\Phi x^h), \Phi:V\to V\otimes W, V$ --
arbitrary highest-weight module then $D=0$.
\endproclaim
\demo{Proof of the lemma} Let us assume that $D\ne 0$. Multiplying
$D$ by a suitable
polynomial of $x_i$ we can assume that $D$ has polynomial
coefficients: $D=\sum x^\alpha D_{(\alpha)}$, $D_{(\alpha)}$ being
difference operators with constant matrix-valued coefficients.  Let us
take the maximal (with respect to the lexicographic ordering) $\alpha$
such that $D_{(\alpha)}\ne 0$. Then if we have a trace $\varphi$ as
above such that $\varphi(x)=x^\lambda w+\text{lower order terms}$ then,
taking the highest term of $D\varphi$,  we see that
$D_{(\alpha)}(x^\lambda w)=0$. On the other hand, if we take $\lambda$
such that $\lambda+\rho\in -P_{+}$ then Verma module $M_\lambda$ is
irreducible and thus for every $w\in W[0]$ there exists a non-zero
intertwiner $\Phi\colon M_\lambda\to M_\lambda\otimes W$ such that the
corresponding trace has the form $\varphi(x)=x^\lambda w+\text{lower
order terms}$. Thus $D_{(\alpha)}(x^\lambda w)=0$ for all $\lambda\in
-P_{+}-\rho, w\in W[0]$. Thus, if one writes $D_{(\alpha)}=\sum_\beta
a_{\alpha\beta}T_\beta$ then $\sum_\beta a_{\alpha\beta}w q^{<\beta,
\lambda>}=0 $ for all $w\in W[0], \lambda\in -P_{+}-\rho$. This is
possible only if all $a_{\alpha\beta}=0$, which contradicts the
assumption $D_{(\alpha)}\ne 0$.\qed
\enddemo \enddemo

In general, $u\mapsto D_u$ is not an algebra homomorphism. However, if
$u$ is central: $u\in \Cal Z(\U)$ then $\Phi u$ is also an intertwiner,
and thus for every $v\in \U$ we have:
$$D_{uv}\Tr(\Phi x^h)=\Tr(\Phi uv x^h)=D_v\Tr(\Phi u x^h)=D_vD_u\Tr (\Phi
x^h).$$
This implies the following proposition:
\proclaim{Proposition 6.1} $u\mapsto D_u$ is an algebra homomorphism
of $\Cal Z(\U)$  to  $\text{DO}\otimes \U[0]/I$, where $I$ is the
ideal generated  by $q^{h_i}-1$. \endproclaim

\proclaim{Proposition 6.2} Let $c\in \Cal Z(\U), V$ be a highest-weight
module over $\U$ (not necessarily finite-dimensional), $c_V=C\Id$ for
some $C\in \C(q)$, and let  $\Phi\colon V\to
V\otimes W$ be a non-zero intertwiner. Then the trace
$\varphi(x)=\Tr|_V(\Phi x^h)$ satisfies the difference equation

$$D_c\varphi(x)=C\varphi(x).\tag 6.3$$
\endproclaim

This proposition is an obvious corollary of Theorem 3.

This shows that our construction allows us to construct  commutative
algebras of difference operators and their eigenfunctions. In general,
these functions are vector-valued (they take values in the space $W[0]$);
however, if we choose $W$ as in Section 4 so that $W[0]$ is
one-dimensional then we can consider the traces as scalar functions;
since every  central element in $\U$ has weight zero, $D_c$
preserves $W[0]$ and thus can be considered as a difference operator with
scalar coefficients. We want to show that for appropriate choice of central
elements the operators $D_c$  are precisely Macdonald's operators (up
to conjugation).

To find these central elements we will use Drinfeld's construction of
central elements (\cite{D2}), which is based on the universal R-matrix
$\Cal R\in \U\hat \otimes \U$ discussed in Section 2 (a similar
construction was independently proposed by N.Reshetikhin \cite{R}).
Define $\Cal
R^{21}=P(\Cal R), P(x\otimes y)=y\otimes x$.

\proclaim{Proposition 6.3}  Define $c_r\in \U, r=1\,\ldots, n$ by
$$c_r=(\text{\rm Id} \otimes \Tr_{(\Lambda^r_q)^*})\left(\Cal R^{21}\Cal R
(1\otimes q^{-2\rho})\right),\tag 6.4$$
where $\Lambda^r_q$ is the $q$-deformation of the representation of $\gl$
in the $r$-th exterior power of the fundamental representation
$\Lambda^r\C^n$.
Then

1. $c_r\in \Cal Z(\U)$

2. If $V$ is a highest-weight module with highest weight $\lambda$,
then
$$c_r|_{V}=\sum\limits_{I}q^{2\sum_{i\in I} (\lambda+\rho)_i}\Id,$$
where the sum is taken over all sets $I=\{i_1,\ldots, i_r\}\subset
\{1,\ldots, n\}$ such that $i_1<\ldots<i_r$.

\endproclaim
\demo{Proof}

1. This is based on the following statement (see \cite{D2}):
if $\theta\colon \U\to \C(q)$ is such that $\theta(xy)=\theta(y S^2(x))$
then the element $c_\theta=(\Id \otimes \theta)(\Cal R^{21}\Cal R)$ is
central. On the other hand, we know that $S^2(x)=q^{-2\rho }x q^{2\rho}$, so
$\theta(x)=\Tr|_V(xq^{-2\rho} )$, where $V$ is any finite-dimensional
representation of $\U$, satisfies $\theta(xy)=\theta(y
S^2(x))$. Taking $V=(\Lambda^r_q)^*$, we get statement 1 of the proposition.

2.  Let $v_\lambda$ be a highest-weight vector in $V$; let us
calculate $c_r v_\lambda$. Let $w\in (\Lambda^r_q)^*[\mu]$. Then (2.7) implies

$$\Cal R^{21}\Cal R (v_\lambda\otimes
w)=q^{-2<\lambda,\mu>}v_\lambda\otimes w + \sum v'_i\otimes w'_i$$
where $\text{wt }w'_i <\mu$. Thus, $c_r v_\lambda =(\sum \limits_\mu
(\text{dim }(\Lambda_q^r)^*[\mu]) q^{-2<\lambda,\mu>}q^{-2<\rho,\mu>})
v_\lambda$,  where the sum is
taken over all the weights of $(\Lambda^r_q)^*$. Since the weights of
$(\Lambda^r_q)^*$ are $\mu=(\mu_1,\ldots,\mu_n)$ such that $\mu_i=0$ or $-1,
\sum \mu_i=-r$, and multiplicity of each weight is 1, we get the desired
formula. \qed\enddemo

\remark{Remark} These central elements are closely related to those
constructed in \cite{FRT}. Essentially, the central elements
constructed in \cite{FRT} are traces of the powers of $L$-matrix,
whereas our central elements are coefficients of the characteristic
polynomial of $L$.\endremark

\proclaim{Theorem 4}
$$M_r=\varphi_0^{-1}(x)\circ D_{c_r}\circ \varphi_0(x),$$
where $M_r$ is Macdonald's operator introduced in Section 1, $c_r$ is
the central element constructed in Proposition 6.3, $\varphi_0$ is the
operator of multiplication by the function $\varphi_0$  defined by
(4.2). \endproclaim

\demo{Proof} This follows from the fact that  $M_r$ and
$\varphi_0^{-1}(x)D_{c_r}\varphi_0(x)$ coincide on the Macdonald's polynomials
$P_\lambda(x)=\varphi_\lambda(x)/\varphi_0(x)$: just compare Proposition 1.1,
Theorem 1 and Proposition 6.3. Repeating the uniqueness arguments
outlined in the proof of Theorem 3, but considering $\lambda\in
P_{+}+(k-1)\rho$ instead of $\lambda\in -P_{+}-\rho$, we see that it
is  only possible if
$M_r=\varphi_0^{-1}\circ D_{c_r}\circ \varphi_0$.
\qed\enddemo

Thus, we can use the traces of the form (4.2) to find eigenfunctions
of Macdonald operators $M_r$. Indeed, let us consider
$\lambda=(\lambda_i,\ldots, \lambda_n)$ as a
formal variable; then $q, q^{\lambda_i}$  are algebraically
independent. In this case Verma module $M_\lambda$ is irreducible, and
thus there exists an intertwiner $\Phi\colon M_\lambda\to
M_\lambda\otimes U$, where the module $U$ is the same we used in
Section 4.

\proclaim{Theorem 5}

1. The function

$$f_\lambda(x)=\frac{\Tr|_{M_\lambda}(\Phi x^h)}{\varphi_0(x)},\tag 6.4$$
where $\Phi:M_\lambda\to M_\lambda\otimes U$ is a non-zero intertwiner
and $\phi_0(x)$ is defined by (4.2),
satisfies the following system of difference equations

$$M_rf_\lambda(x)=\sum\limits_Iq^{2\sum_{i\in I}
(\lambda+\rho)_i}f_{\lambda}(x) \tag 6.5$$

2. The functions $f_{\sigma(\lambda+\rho)-\rho}, \sigma\in S_n$ form a basis of
solutions of the  system (6.5) in the space of generalized Laurent
series
$\Cal F=\sum\limits_\nu x^\nu\C(q,
q^{\lambda_i})[[\frac{x_2}{x_1},\ldots, \frac{x_n}{x_{n-1}}]]$
\endproclaim
\demo{Proof}

1. This is an immediate corollary of Proposition 6.2 and
Theorem 4.

2. Suppose that $f\in \Cal F$ is a solution of (6.5) of the form

$$f(x)=x^\nu+\text {lower order terms}.$$

Expanding coefficients of Macdonald's operators in Laurent series, we
find the highest term of $M_r f$:

$$(M_rf)(x)=
\sum\limits_{I: |I|=r} q^{2\sum_{i\in I}\rho_i}T_{q^2,x_I}f
=\sum\limits_{I:|I|=r}q^{2\sum_{i\in
I}(\nu+\rho)_i}x^\nu+\ldots$$

Thus $f(x)$ can be a solution only if for any $r$,
$$\sum\limits_{I:|I|=r}q^{2\sum_{i\in
I}(\nu+\rho)_i}=\sum\limits_{I:|I|=r}q^{2\sum_{i\in I}(\lambda
+\rho)_i},$$
which is only possible if $\nu+\rho=\sigma(\lambda+\rho)$ for some
$\sigma\in S_n$. Then the highest term of $f$ coincides with the
highest term of $f_{\sigma(\lambda+\rho)-\rho}$. Considering
$f-f_{\sigma(\lambda+\rho)-\rho}$ and repeating the same arguments,
we finally see that $f$ is a linear combinatin of the functions
$f_{\sigma(\lambda+\rho)-\rho}$.
\qed\enddemo

\enddocument
\end